# On Optimal Linear Redistribution of VCG Payments in Assignment of Heterogeneous Objects

Sujit Gujar and Y Narahari

Dept of Computer Science and Automation
Indian Institute of Science, Bangalore,560012.
{sujit,hari}@csa.iisc.ernet.in

**Abstract.** There are $p$ heterogeneous objects to be assigned to $n$ competing agents ($n > p$) each with unit demand. It is required to design a Groves mechanism for this assignment problem satisfying weak budget balance, individual rationality, and minimizing the budget imbalance. This calls for designing an appropriate rebate function. Our main result is an impossibility theorem which rules out linear rebate functions with non-zero efficiency in heterogeneous object assignment. Motivated by this theorem, we explore two approaches to get around this impossibility. In the first approach, we show that linear rebate functions with non-zero are possible when the valuations for the objects are correlated. In the second approach, we show that rebate functions with non-zero efficiency are possible if linearity is relaxed.

## 1 Introduction

Consider that $p$ public resources are available and $n > p$ agents are interested in utilizing one of them. Naturally. we should assign these resource such that those who value them most get it. Since Groves mechanisms [13,3,6] have attractive game theoretic properties such as dominant strategy incentive compatibility (DSIC) and allocative efficiency (AE), Groves mechanisms are widely used in practice. However, in general, a Groves mechanism need not be budget balanced. That is, the total transfer of money in the system may not be zero. So the system will be left with a surplus or deficit. Using Clarke's mechanism [3], we can ensure under fairly weak conditions that there is no deficit of money, that is the mechanism is weakly budget balanced. In such a case, the system or the auctioneer will be left with some money.

Often, the surplus money is not really needed in many social settings such as allocations by the Government among its departments, etc. Since strict budget balance cannot coexist with DSIC and AE (Green-Laffont theorem [5]), we would like to redistribute the surplus to the participants as far as possible, preserving DSIC and AE. This idea was originally proposed by Laffont [11]. The total payment made by the mechanism as a redistribution will be referred to as the *rebate* to the agents.

In this paper, we consider the following problem. There are $n$ agents and $p$ heterogeneous objects ($n \geq p > 1$). Each agent desires exactly one object out of these $p$ objects. His valuation for any of the objects is independent of his valuations for the other objects. Valuations of the different agents are also mutually independent. Our goal is to design a mechanism for assignment of the $p$ objects among the $n$ agents which is allocatively efficient, dominant strategy incentive compatible, and maximizes the rebate (which is equivalent to minimizing the budget imbalance). In addition, we would like the mechanism to satisfy feasibility and individual rationality. Thus, we seek to design a Groves mechanism for assigning $p$ heterogeneous objects among $n$ agents satisfying:

1. Feasibility (F) or weak budget balance. That is, the total payment to the agents should be less than or equal to the total received payment.
2. Individual Rationality (IR), which means that each agent's utility by participating in the mechanism
   should be non-negative.
3. Minimizes budget imbalance.

We call such a mechanism *Groves redistribution mechanism* or simply *redistribution mechanism*. Designing a redistribution mechanism involves design of an appropriate *rebate function*. If in a redistribution mechanism, the rebate function for each agent is a linear function of the valuations of the remaining agents, we refer to such mechanism as a *linear redistribution mechanism*, (LRM). In many situations, design of an appropriate LRM turns out to be a problem of solving a linear program.

Due to the Green-Laffont theorem [5], we cannot guarantee 100% redistribution at all type profiles. So a performance index for the redistribution mechanism would be the worst case redistribution. That is, the fraction of the surplus which is guaranteed to be redistributed irrespective of the bid profiles. This fraction will be referred to as *efficiency* in the rest of the paper (*Note*: This efficiency is not to be confused with allocative efficiency). The advantage of worst case analysis is that, it does not require any distributional information on the type sets of the agents. It is desirable that the rebate function is deterministic and anonymous. A rebate function is said to be anonymous if two agents having the same bids get the same rebate. So, the aim is to design an anonymous, deterministic rebate function which maximizes the efficiency and satisfies feasibility and individual rationality.

Our paper seeks to extend the results of Moulin [12] and Guo and Conitzer [8] who have independently designed a Groves mechanism in order to redistribute the surplus when objects are identical (homogeneous objects case). Their mechanism is deterministic, anonymous, and has maximum efficiency over all possible Groves redistribution mechanisms. We will refer to their mechanism as the *worst case optimal* (WCO) mechanism or *Moulin* mechanism. WCO/Moulin Mechanisms are linear redistribution mechanisms. In this paper, we concentrate on designing a linear redistribution mechanism for the heterogeneous objects case.

### 1.1   Relevant Work

As it is impossible to achieve allocative efficiency, DSIC, and budget balance simultaneously, we have to compromise on either of these properties. Faltings [4] and Guo and Conitzer [9] achieve budget balance by compromising on AE. If we are interested in preserving AE and DSIC, we have to settle for a non-zero surplus or a non-zero deficit of the money (budget imbalance) in the system. To reduce budget imbalance, various rebate functions have been designed by Bailey [1], Cavallo [2], Moulin [12], and Guo and Conitzer [8]. Moulin [12] and Guo and Conitzer [8] designed a Groves redistribution mechanism for assignments of $p$ homogeneous objects among $n > p$ agents with unit demand. Guo and Conitzer [10] designed a redistribution mechanism which is optimal in the expected sense for the homogeneous objects setting. Thus, it will require some distributional information over the type sets of the agents. Sujit and Narahari [7] have designed a redistribution mechanism for the assignments of $p$ heterogeneous objects among $n$ competing agents with unit demands. The rebate function proposed by them is not a linear function.

### 1.2   Contributions and Outline

In this paper, we investigate the question of existence of a linear rebate function for redistribution of surplus in assignment of heterogeneous objects. Our result shows that in general, when the domain of valuations for each agent is $\mathbb{R}_+^p$, it is impossible to design a linear rebate function, with non-zero efficiency, for the heterogeneous settings. However, we can relax the assumption of independence of valuations of different objects to get a linear rebate function with non-zero efficiency. Another way to get around the impossibility theorem is to relax the linearity requirement of a rebate function. In particular, our contributions in this paper can be summarized as follows.

- We first prove the impossibility of existence of a linear rebate function with non-zero efficiency for the heterogeneous settings when the domain of valuations for each agent is $\mathbb{R}_+^p$ and the valuations for the objects are independent.
- When the objects are heterogeneous but the values for the objects of an agent can be derived from one single number, that is, the private information is still single dimensional, we design a Groves redistribution mechanism which is linear, anonymous, deterministic, feasible, individually rational, and efficient. In addition, the mechanism is worst case optimal.
- We show the existence of a non-linear rebate function that has non-zero efficiency.

The paper is organized as follows. In Section 2, we introduce the notation followed in the paper and describe some background work from the literature. In Section 3, we state and prove the impossibility result. We derive an extension of Moulin/ WCO mechanism for heterogeneous objects but with single dimensional private information in Section 4. The impossibility result does not rule out possibility of non-linear rebate functions with strictly positive efficiency. We show this with a redistribution mechanism, BAILEY, which is Bailey's mechanism [1] applied to the settings under consideration in Section 5. We will conclude the paper in Section 6.

We need an ordering of the bids of the agents which we define in Appendix A.

## 2  Preliminaries and Notation

**Table 1.** Notation

| | |
|---|---|
| $n$ | Number of agents |
| $N$ | Set of the agents = $\{1, 2, \ldots, n\}$ |
| $p$ | Number of objects |
| $i$ | Index for an agent, $i = 1, 2, \ldots, n$ |
| $j$ | Index for object, $j = 1, 2, \ldots, p$ |
| $\mathbb{R}_+$ | Set of positive real numbers |
| $\theta_{ij}$ | Valuation of an agent $i$ for an object $j$ |
| $\theta_i$ | A vector of valuations of the agent $i$, $= (\theta_{i1}, \theta_{i2}, \ldots, \theta_{ip})$ |
| $\Theta_i$ | The space of valuations of agent $i$, $= \mathbb{R}_+^p$ |
| $\Theta$ | $\Theta_1 \times \Theta_2 \ldots \times \Theta_n$ |
| $b_i$ | Bid submitted by agent $i$, $= (b_{i1}, b_{i2}, \ldots, b_{ip}) \in \Theta_i$ |
| $b$ | $(b_1, b_2, \ldots, b_n)$, the bid vector |
| $K$ | The set of all allocations of $p$ objects to $n$ agents, each getting at most one object |
| $k(b)$ | An allocation, $k(.) \in K$, corresponding to the bid profile b |
| $k^*(b)$ | An allocatively efficient allocation when the bid profile is $b$ |
| $k^*_{-i}(b)$ | An allocatively efficient allocation when the bid profile is $b$ and agent $i$ is excluded from the system |
| $v_i(k(b))$ | Valuation of the allocation $k$ to the agent $i$, when $b$ is the bid profile |
| $v$ | $v : K \to \mathbb{R}$, the valuation function, $v(k(b)) = \sum_{i \in N} v_i(k(b))$ |
| $t_i(b)$ | Payment made by agent $i$ in the Clarke's pivotal mechanism, when the bid profile is $b$ $t_i(b) = v_i(k^*(b)) - \left(v(k^*(b)) - v(k^*_{-i}(b))\right)$ |
| $t(b)$ | The Clarke payment, that is, the total payment received from all the agents, $t(b) = \sum_{i \in N} t_i$ |
| $t^{-i}$ | The Clarke payment received in the absence of the agent $i$ |
| $r_i(b)$ | Rebate to agent $i$ when bid profile is $b$ |
| $e$ | The efficiency of the mechanism, $= \inf_{\theta : t \neq 0} \frac{\sum r_i(\theta)}{t(\theta)}$ |

The notation used is summarized in Table 1. Note that, where the context is clear, we will use $t, t_i, r_i, k,$ and $v_i$ to indicate $t(b), t_i(b), r_i(b), k(b),$ and $v_i(k(b))$ respectively. In this paper, we assume that the payment made by agent $i$ is of the form $t_i(\cdot) - r_i(\cdot)$, where $t_i(\cdot)$ is agent $i$'s payment in the Clarke's pivotal mechanism [3]. We refer to $\sum_i t_i$, as the total Clarke's payment or the surplus in the system.

### 2.1  Optimal Worst Case Redistribution when Objects are Identical

When the objects are identical, every agent $i$ has the same value for each object, call it $v_i$. Without loss of generality, we will assume, $v_1 \geq v_2 \geq \ldots \geq v_n$. In Clarke's pivotal mechanism, the first $p$ agents will receive the objects and each of these $p$ agents will pay $v_{p+1}$. So, the surplus in the system is $pv_{p+1}$. For this situation, Moulin [12] and Guo and Conitzer [8] have independently designed a redistribution mechanism.

Guo and Conitzer [8] maximize the worst case fraction of total surplus which gets redistributed. This mechanism is called the WCO mechanism. Moulin [12] minimizes the ratio of budget imbalance to the value of an optimal allocation, that is the value of an allocatively efficient allocation. The WCO mechanism coincides with Moulin's feasible and individually rational mechanism. Both the above mechanisms work as follows. After receiving bids from the agents, bids are sorted in decreasing order. The first $p$ agents receive the objects. Each agent's Clarke's payment is calculated, say $t_i$. Every agent $i$ pays, $p_i = t_i - r_i$, where, $r_i$ is the rebate function for an agent $i$. Suppose $y_1 \geq y_2 \geq \ldots \geq y_{n-1}$ are the bids of the $(n-1)$ agents excluding the agent $i$, then the rebate to the agent $i$ is given by,

$$r_i^{WCO} = \sum_{j=p+1}^{n-1} c_j y_j \tag{1}$$

where,

$$c_j = \frac{(-1)^{j+p-1}(n-p)\binom{n-1}{p-1}}{j\binom{n-1}{j}\sum_{k=p}^{n-1}\binom{n-1}{k}} \left\{ \sum_{k=j}^{n-1} \binom{n-1}{k} \right\} \tag{2}$$

for   $i = p+1, \ldots, n-1$.

The efficiency of this mechanism is $e^*$, where $e^*$ is given by,

$$e^* = 1 - \frac{\binom{n-1}{p}}{\sum_{k=p}^{n-1}\binom{n-1}{k}}$$

This has been shown to be optimal in sense that no other mechanism can guarantee greater than $e^*$ fraction redistribution in the worst case.

## 3  Impossibility of Linear Rebate Function with Non-Zero Efficiency

We have just reviewed the design of a redistribution mechanism for homogeneous objects. We have seen that the Moulin/WCO mechanism is a linear function of the types of agents. We now explore the general case. In the homogeneous case, the bids are real numbers which can be arranged in decreasing order. The Clarke's surplus is a linear function of these ordered bids. For the heterogeneous scenario, this would not be the case. Each bid $b_i$ belongs to $\mathbb{R}_+^p$; hence, there is no unique way of defining an order among the bids. Moreover, the Clarke's surplus is not a linear function of received bids in the heterogeneous case. So, we cannot expect any linear/affine rebate function of types to work well at all type profiles. We will prove this formally.

We first generalize a theorem due to Guo and Conitzer [8]. The context in which Guo and Conitzer [8] stated and proved the theorem is in the homogeneous setting. We now show that this result holds true in the heterogeneous objects case also. The symbol $\succcurlyeq$ denotes the order over the bids of the agents, as defined in the Appendix A.2.

**Theorem 1.** *Any deterministic, anonymous rebate function $f$ is DSIC iff,*

$$r_i = f(v_1, v_2, \ldots, v_{i-1}, v_{i+1}, \ldots, v_n) \ \forall i \in N \tag{3}$$

where, $v_1 \succcurlyeq v_2 \succcurlyeq \ldots \succcurlyeq v_n$.

**Proof:** We provide only a sketch of the proof.

– The "if" part: If $r_i$ takes the form given by equation (3), then the rebate of agent $i$ is independent of his valuation. The allocation rule satisfies allocative efficiency. So, the mechanism is still Groves and hence DSIC. The rebate function defined is deterministic. If two agents have the same bids, then, as per the ordering defined in Appendix, $\succcurlyeq$, they will have the same ranking. Suppose agents $i$ and $i+1$ have the same bids. Thus $v_i \succcurlyeq v_{i+1}$ and $v_{i+1} \succcurlyeq v_i$. So, $r_i = f(v_1, v_2, \ldots, v_{i-1}, v_{i+1}, \ldots, v_n)$ and $r_{i+1} =$
$f(v_1, v_2, \ldots, v_i, v_{i+2}, \ldots, v_n)$. Since $v_i = v_{i+1}$, $r_i = r_{i+1}$. Thus the rebate function is anonymous.

– The "only if" part: The homogeneous objects case is a special case of the mechanism. When objects are homogeneous, the ordering of the bids $\succcurlyeq$ matches the $\geq$ ordering on real numbers. If the rebate function is not in the form defined in the theorem, the rebate function would not simultaneously satisfy the DSIC, anonymity, and deterministic properties. This is because the above form of the rebate function is a necessary condition when the objects are identical. Thus we need a rebate function in this form in heterogeneous settings as well.

∎

We now state and prove the main result of this paper.

**Theorem 2.** *If a redistribution mechanism is feasible and individually rational, then there cannot exist a linear rebate function which satisfies all the following properties:*

- *DSIC*
- *deterministic*
- *anonymous*
- *non-zero efficiency.*

**Proof :** Assume that there exists a linear function, say $f$, which satisfies the above properties. Let $v_1 \succcurlyeq v_2 \succcurlyeq \ldots \succcurlyeq v_n$. Then according to Theorem 1, for each agent $i$,

$$r_i = f(v_1, v_2, \ldots, v_{i-1}, v_{i+1}, \ldots, v_n)$$
$$= (c_0, e_p) + (c_1, v_1) + \ldots + (c_{n-1}, v_n)$$

where, $c_i = (c_{i1}, c_{i2}, \ldots, c_{ip}) \in \mathbb{R}^p$, $e_p = (1, 1, \ldots, 1) \in \mathbb{R}^p$, and $(\cdot, \cdot)$ denotes the inner product of two vectors in $\mathbb{R}^p$. Now, we will show that the worst case performance of $f$ will be zero. To this end, we will study the structure of $f$, step by step.

<u>Observation 1:</u> Consider type profile $(v_1, v_2, \ldots, v_n)$ where $v_1 = v_2 = \ldots = v_n = (0, 0, \ldots, 0)$. For this type profile, the total Clarke's surplus is zero and $r_i = (c_0, e_p) \,\forall i \in N$. Individual rationality implies,

$$(c_0, e_p) \geq 0 \qquad (4)$$

Feasibility should imply the total redistributed amount is less than the surplus, that is,

$$\sum_i r_i = n(c_0, e_p) \leqslant 0 \qquad (5)$$

From, (4) and (5), it is easy to see that, $(c_0, e_p) = 0$.

<u>Observation 2:</u> Consider type profile $(v_1, v_2, \ldots, v_n)$ where $v_1 = (1, 0, 0, \ldots, 0)$ and $v_2 = \ldots, v_n = (0, 0, \ldots, 0)$. For this type profile, $r_1 = 0$ and if $i \neq 1$, $r_i = c_{11} \geq 0$ for individual rationality. For this type profile, the Clarke's surplus is zero. Thus, for feasibility, $\sum_i r_i = (n-1)c_{11} \leq t = 0$. This implies, $c_{11} = 0$.

In the above profile, by considering $v_1 = (0, 1, , 0, \ldots, 0)$, we get $c_{12} = 0$. Similarly, one can show $c_{13} = c_{14} = \ldots = c_{1p} = 0$.

<u>Observation 3:</u> Continuing like above with, $v_1 = v_2 = \ldots = v_i = e_p$, and $v_{i+1} = (1, 0 \ldots, 0)$ or $(0, 1, 0 \ldots, 0), \ldots$ or $(0, \ldots, 0, 1)$, we get, $c_{i+1} = (0, 0, \ldots, 0) \,\forall\, i \leq p-1$. Thus,

$$r_i = \begin{cases} (c_{p+1}, v_{p+2}) + \ldots + (c_{n-1}, v_n) & : \text{if } i \leq p+1 \\ (c_{p+1}, v_{p+1}) + \ldots + (c_{i-1}, v_{i-1}) \\ \quad + (c_i, v_{i+1}) + \ldots + (c_{n-1}, v_n) & : \text{otherwise} \end{cases}$$

We now claim that efficiency of this mechanism is zero. That is, in the worst case, the fraction of the Clarke's surplus that gets redistributed is zero. Suppose we show that there exists a type profile, for which the Clarke's surplus non-zero and the rebate to each agent is zero. Then the theorem is proved. So, it remains to show the existence of such a type profile. Consider the type profile:

$$\begin{aligned} v_1 &= (2p-1, 2p-2, \ldots, p) \\ v_2 &= (2p-2, 2p-3, \ldots, p-1) \\ &\vdots \\ v_p &= (p, p-1, \ldots, 1) \end{aligned}$$

and $v_{p+1} = v_{p+2} \ldots = v_n = (0, 0, \ldots, 0)$.

Now, with this type profile, agent 1 pays $(p-1)$, agent 2 pays $(p-2), \ldots$, agent $(p-1)$ pays 1 and the remaining agents pay 0. Thus, the Clarke's payment received is non-zero but it can be seen that $r_i = 0$ for all agents.

∎

The above theorem provides a disappointing news. It rules out the possibility of a linear redistribution mechanism for heterogeneous settings which will have non-zero efficiency. However, there are two ways to get around it.

1. The domain of types under which Theorem 2 holds is, $\Theta_i = \mathbb{R}_+^p$, $\forall\, i \in N$. One idea is to restrict the domain of types. In Section 3, we design a worst case optimal linear redistribution mechanism when the valuations of agents for the heterogeneous objects have a certain type of correlation.
2. Explore the existence of a rebate function which is not a linear and yields a non-zero performance. We explore this in Section 5.

## 4  A Redistribution Mechanism for Heterogeneous Objects having Scaling Based Correlation

Consider a scenario where the objects are not identical but the valuations for the objects are correlated and can be derived by a single parameter. As a motivating example, consider the website `somefreeads.com` and assume that there are $p$ slots available for advertisements and there are $n$ agents interested in displaying their ads. Naturally, every agent will have a higher preference for a higher slot. Define *click through rate* of a slot as the number of times the ad is clicked, when the ad is displayed in that slot, divided by the number of impressions. Let the click through rates for slots be $\alpha_1 \geq \alpha_2 \geq \alpha_3 \ldots \geq \alpha_p$. Assume that each agent has the same value for each click by the user, say $v_i$. So, the agent's value for the $j^{th}$ slot will be $\alpha_j v_i$. Let us use the phrase *valuations with scaling based correlation* to describe such valuations. We define this more formally below.

**Definition 1.** *We say the valuations of the agents have scaling based correlation if there exist positive real numbers $\alpha_1, \alpha_2, \alpha_3, \ldots, \alpha_p > 0$ such that, for each agent $i \in N$, the valuation for object $j$ is of the form $\theta_{ij} = \alpha_j v_i$, where $v_i \in \mathbb{R}_+$ is a private signal observed by agent $i$.*

Without loss of generality, we assume, $\alpha_1 \geq \alpha_2 \geq \alpha_3 \ldots \geq \alpha_p > 0$.

For the above setting, we design a Groves mechanism which is almost budget balanced and optimal in the worst case. Our mechanism is similar to that of Guo and Conitzer [8] and our proof uses the same line of arguments.

The following theorem by Guo and Conitzer [8] will be used to design our mechanism.

**Theorem 3.** *For any $x_1 \geq x_2 \geq \ldots x_n \geq 0$,*

$$a_1 x_1 + a_2 x_2 + \ldots a_n x_n \geq 0 \ \text{ iff } \ \sum_{i=1}^{j} a_i \geq 0 \ \ \forall j = 1, 2 \ldots, n$$

### 4.1 The Proposed Mechanism

We will use a linear rebate function. (For making equations symmetric, we will assume that there are $(n-p)$ virtual objects, with $\alpha_{p+1} = \alpha_{p+2} = \ldots = \alpha_n = 0$). We propose the following mechanism:

- The agents submit their bids.
- Their bids are sorted in decreasing order.
- The highest bidder will be allotted the first object, the second highest bidder will be allotted the second object, and so on.
- Agent $i$ will pay $t_i - r_i$, where $t_i$ is the Clarke's payment and $r_i$ is the rebate.

$$t_i = \sum_{j=i}^{p}(\alpha_j - \alpha_{j+1})v_{j+1}$$

- Let agent $i$'s rebate be,

$$r_i = c_0 + c_1 v_1 + \ldots + c_{i-1}v_{i-1} + c_i v_{i+1} + \ldots + c_{n-1}v_n$$

The mechanism is required to be individually rational and feasible.

- The mechanism will be individually rational iff $r_i \geq 0 \; \forall i \in N$. That is, $\forall \, i \in N$,

$$c_0 + c_1 v_1 + \ldots + c_{i-1}v_{i-1} + c_i v_{i+1} + \ldots + c_{n-1}v_n \geq 0.$$

- The mechanism will be feasible if the total redistributed payment is less than or equal to the surplus. That is, $\sum_i r_i \leq t = \sum_i t_i$ or $t - \sum_i r_i \geq 0$, where,

$$t = \sum_{j=1}^{p} j(\alpha_j - \alpha_{j+1})v_{j+1}.$$

With the above setup, we now derive $c_0, c_1, \ldots, c_{n-1}$ that will maximize the fraction of surplus which is redistributed among the agents.

Step 1: First, we claim that, $c_0 = c_1 = 0$. This can be proved as follows. Consider the type profile, $v_1 = v_2 = \ldots = v_n = 0$. For this type profile, individual rationality implies $r_i = c_0 \geq 0$ and $t = 0$. So for feasibility, $\sum_i r_i = nc_0 \leq t = 0$. That is, $c_0$ should be zero. Similarly, by considering type profile $v_1 = 1, v_2 = \ldots = v_n = 0$, we get $c_1 = 0$.

∎

Step 2: Using $c_0 = c_1 = 0$,

- The feasibility condition can be written as:

$$\sum_{j=2}^{n-1}\Big((j-1)(\alpha_{j-1} - \alpha_j) - (j-1)c_{j-1} - (n-j)c_j\Big)v_j - (n-1)c_{n-1}v_n \geq 0 \quad (6)$$

- The individual rationality condition can be written as

$$c_2 v_2 + \ldots + c_{i-1}v_{i-1} + c_i v_{i+1} + \ldots + c_{n-1}v_n \geq 0 \quad (7)$$

Step 3: When we say our mechanism's efficiency is $e$, we mean, $\sum_i r_i \geq et$, that is,

$$\sum_{j=2}^{n-1}\Big(-e(j-1)(\alpha_{j-1} - \alpha_j) + (j-1)c_{j-1} + (n-j)c_j\Big)v_j + (n-1)c_{n-1}v_n \geq 0 \quad (8)$$

<u>Step 4</u>: Define $\beta_1 = \alpha_1 - \alpha_2$, and for $i = 2, \ldots, p$, let $\beta_i = i(\alpha_i - \alpha_{i+1}) + \beta_{i-1}$. Now, inequalities (6), (7), and (8) have to be satisfied for all values of $v_1 \geq v_2 \geq \ldots \geq v_n$. Invoking Theorem (3), we need to satisfy the following set of inequalities:

$$\sum_{i=2}^{j} c_i \geq 0 \ \ \forall j = 2, \ldots n-1$$
$$e\beta_1 \leq (n-2)c_2 \leq \beta_1$$
$$e\beta_{i-1} \leq n \sum_{j=2}^{i-1} c_j + (n-i)c_i \leq \beta_{i-1} \ \ i = 3, \ldots, p$$
$$e\beta_p \leq n \sum_{j=2}^{i-1} c_j + (n-i)c_i \leq \beta_p \ \ i = p+1, \ldots, n-1$$
$$e\beta_p \leq n \sum_{j=2}^{n-1} c_j \leq \beta_p$$

Now, the social planner wishes to design a mechanism that maximizes $e$ subject to the above constraints.

Define $x_j = \sum_{i=2}^{j} c_i$ for $j = 2, \ldots, n-1$. This is equivalent to solving the following linear program.

$$\boxed{\begin{array}{c} \text{maximize } e \\ \text{s.t.} \\ e\beta_1 \leq (n-2)x_2 \leq \beta_1 \\ e\beta_{i-1} \leq ix_{i-1} + (n-i)x_i \leq \beta_{i-1} \ \ i = 3, \ldots, p \\ e\beta_p \leq ix_{i-1} + (n-i)x_i \leq \beta_p \ \ i = p+1, \ldots, n-1 \\ e\beta_p \leq nx_{n-1} \leq \beta_p \\ x_i \geq 0 \ \ \forall i = 2, \ldots, n-1 \end{array}} \quad (9)$$

So, given $n$ and $p$, the social planner will have to solve the above optimization problem and determine the optimal values of $e, c_2, c_3, \ldots, c_{n-1}$.

The discussion above can be summarized as the following theorem.

**Theorem 4.** *When the valuations of the agents have scaling based correlation, for any $p$ and $n > p+1$, the linear redistribution mechanism obtained by solving LP (9) is worst case optimal among all Groves redistribution mechanisms that are feasible, individually rational, deterministic, and anonymous.*

**Proof:** This can be proved following the line of arguments of Guo and Conitzer [8].

■

## 5 Non-linear Redistribution Mechanisms for the Heterogeneous Setting

We should note that the homogeneous objects case is a special case of the heterogeneous objects case in which each bidder submits the same bid for all objects. Thus, we cannot expect any redistribution mechanism to perform better than the homogeneous objects case. For $n \leq p + 1$, the worst case redistribution is zero for the homogeneous case and so will be for the heterogeneous case. So, we assume $n > p + 1$. We propose a redistribution mechanism which we will refer to as *BAILEY* mechanism. It is to be noted that the BAILEY redistribution scheme the mechanism proposed by Bailey [1] applied to the heterogeneous setting.

### 5.1 BAILEY Mechanism

First, consider the case when $p = 1$. Let the valuations of the agents for the object be, $v_1 \geq v_2 \geq \ldots \geq v_n$. The agent with the highest valuation will receive the object and would pay the second highest bid. Cavallo [2] proposed the rebate function as,

$$r_1 = r_2 = \tfrac{1}{n} v_3$$
$$r_i = \tfrac{1}{n} v_2 \ \ i > 2$$

Motivated by this scheme, we propose a scheme for the heterogeneous setting. Suppose agent $i$ is excluded from the system. Then let $t^{-i}$ be the Clarke's surplus in the system (defined in Table 1). Define,

$$r_i{}^B = \frac{1}{n} t^{-i} \ \ \forall \ i \ \in \ N \quad (10)$$

- As the Clarke surplus is always positive, $r_i{}^B \geq 0$ for all $i$. Thus, this scheme satisfies individual rationality.
- $t^{-i} \leq t \ \forall \ i$. So, $\sum_i r_i{}^B = \sum_i \frac{1}{n} t^{-i} \leq n \frac{1}{n} t = t$. Thus, this scheme is feasible.

We now show that the BAILEY scheme has non-zero efficiency if $n \geq 2p + 1$. First we prove two lemmas. These lemma's are useful in designing redistribution mechanisms for the heterogeneous settings as well as in analysis of the mechanisms. Lemma 2 is used to show non-zero efficiency of the BAILEY mechanism. Lemma 1 is used to find allocatively efficient outcome for the settings under consideration. Also, this lemma 1 is useful in determining Clarke's payments.

**Lemma 1.** *If we sort the bids of all the agents for each object, then*

1. *An optimal allocation, that is an allocatively efficient allocation, will consist of the agents having bids among the $p$ highest bids for each object.*
2. *If any one of the agents from an optimal allocation is removed from the system, there exists an optimal allocation in which the remaining $(p-1)$ agents receive the objects, perhaps not the same objects as in the original optimal allocation.*

**Proof:**

- Suppose an optimal allocation contains an agent whose bid for his winning object, say $j$, is not in the top $p$ bids for the $j^{th}$ object. There are other $(p-1)$ winners in an optimal allocation. So, there exists at least one agent whose bid is in the top $p$ bids for the $j^{th}$ object and does not win any object. Thus, allocating him the $j^{th}$ object, we have an allocation which has higher valuation than the declared optimal allocation.
- Suppose an agent $i$ who receives an object in an optimal allocation is removed from the system. The agent will have at most one bid in the top $p$ bids for each object. So, agents now having bids in the top $p$ bids, will be at the $p^{th}$ position. It can be seen that there will be at most one agent in an optimal allocation who is on the $p^{th}$ position for the object he wins. If there is more than one agent in an optimal allocation on the $p^{th}$ position for the object they win, then we can improve on this allocation. Hence, after removing $i$, there will be at most one more agent who will be a part of a new optimal allocation.

∎

**Lemma 2.** *There are at most $2p$ agents involved in deciding the Clarke's payment.*

**Proof:** The argument is as follows:

1. Sort the bids of the agents for each object.
2. The optimal allocation consists of agents having bids in the $p$ highest bids for each of the objects (Lemma 1).
3. For computing the Clarke's payment of the agent $i$, we remove the agent and determine an optimal allocation. And, using his bid, the valuation of optimal allocation with him and without him will determine his payment. This is done for each agent $i$. As per Lemma 1, if any agent from an optimal allocation is removed from the system, there exists a new optimal allocation which consists of at least $(p-1)$ agents who received the objects in the original optimal allocation.
4. There will be $p$ agents receiving the objects and determining their payments will involve removing one of them at a time, there will be at most $p$ more agents who will influence the payment. Thus, there are at most $2p$ agents involved in determining the Clarke's payment.

∎

*Note*: When the objects are identical, the bids of $(p+1)$ agents are involved in determining the Clarke's payment.

Now, we show non-zero efficiency of the BAILEY redistribution scheme.

**Proposition 1.** *The BAILEY redistribution scheme has non-zero efficiency, if $n \geq 2p + 1$.*

**Proof:** In Lemma 2 (See Appendix), we have shown that there will be at most $2p$ agents involved in determining the Clarke surplus. Thus, given a type profile, there will be $(n-2p)$ agents, for whom, $t^{-i} = t$ and this implies that at least $\frac{n-2p}{n}t$ will be redistributed.

∎

Note: The proof of Proposition 1 indicates that the efficiency of this mechanism is at least $\frac{n-2p}{n}$. Before we conclude, we would like to state an algorithmic implication of Lemma 1[1].

**Algorithmic Implication of Lemma 1**

To implement the Clarke's mechanism, a naive approach will be to consider all possible feasible allocations and find out an optimal allocation. There will be $\mathcal{C} = n(n-1)\ldots(n-p+1)$ feasible allocations. An optimal allocation will be found in $O(\mathcal{C})$ time. For calculating payments for each agent in an optimal allocation, we have to find an optimal allocation without that agent, which will amount to $O(p * \mathcal{C}')$ time complexity, where $\mathcal{C}' = (n-1)(n-2)\ldots(n-p)$. So total time complexity will be $O(p * \mathcal{C}' + \mathcal{C})$.

Now as per lemma 1, for finding an optimal allocation, we can concentrate only on the top $p$ bids for each object. This can be done in $O(pn \log n)$ time and now the number of allocations which we have to consider, will be $\mathcal{C}'' = p(p-1)\ldots(1) = p!$. So we can determine the VCG payment in $O((p+1)\mathcal{C}'' + pn \log n)$ time complexity. This will be much faster when $n$ is very large compared to $p$.

∎

## 6  Conclusion

We addressed the problem of assigning $p$ heterogeneous objects among $n > p$ competing agents. When the valuations of the agents are independent of each other and their valuations for each object are independent of valuations on the other objects, we proved the impossibility of the existence of a linear redistribution mechanism with non-zero efficiency. Then we explored two approaches to get around this impossibility. In the first approach, we showed that linear rebate functions with non-zero are possible when the valuations for the objects have scaling based correlation. In the second approach, we showed that rebate functions with non-zero efficiency are possible if linearity is relaxed.

It would be interesting to see if we can characterize the situations under which linear redistribution mechanisms with non-zero efficiency are possible for heterogeneous settings. Another interesting problem to explore is to design redistribution mechanisms that are worst case optimal for heterogeneous settings.

## References

1. Martin J Bailey. The demand revealing process: To distribute the surplus. *Public Choice*, 91(2):107–26, April 1997.
2. Ruggiero Cavallo. Optimal decision-making with minimal waste: strategyproof redistribution of VCG payments. In *AAMAS '06: Proceedings of the Fifth International Joint Conference on Autonomous Agents and Multiagent Systems*, pages 882–889, New York, NY, USA, 2006. ACM.
3. E. Clarke. Multi-part pricing of public goods. *Public Choice*, 11:17–23, 1971.
4. Boi Faltings. A budget-balanced, incentive-compatible scheme for social choice. In *Agent-Mediated Electronic Commerce, AMEC*, pages 30–43. Springer, 2005.
5. J. R. Green and J. J. Laffont. *Incentives in Public Decision Making*. North-Holland Publishing Company, Amsterdam, 1979.
6. T. Groves. Incentives in teams. *Econometrica*, 41:617–631, 1973.
7. Sujit Gujar and Yadati Narahari. Redistribution of VCG payments in assignment of heterogeneous objects. In Christos H. Papadimitriou and Shuzhong Zhang, editors, *WINE*, volume 5385 of *Lecture Notes in Computer Science*, pages 438–445. Springer, 2008.

---

[1] Authors like to include this here, since Lemma 1 can be used to design algorithms for conducting redistribution mechanisms in heterogeneous settings

8. Mingyu Guo and Vincent Conitzer. Worst-case optimal redistribution of VCG payments. In *EC '07: Proceedings of the 8th ACM conference on Electronic Commerce*, pages 30–39, New York, NY, USA, 2007. ACM.
9. Mingyu Guo and Vincent Conitzer. Better redistribution with inefficient allocation in multi-unit auctions with unit demand. In *EC '08: Proceedings of the 9th ACM conference on Electronic commerce*, pages 210–219, New York, NY, USA, 2008. ACM.
10. Mingyu Guo and Vincent Conitzer. Optimal-in-expectation redistribution mechanisms. In *AAMAS '08: Proceedings of the 7th international joint conference on Autonomous agents and multiagent systems*, pages 1047–1054, Richland, SC, 2008. International Foundation for Autonomous Agents and Multiagent Systems.
11. J.J. Laffont and E. Maskin. A differential approach to expected utility maximizing mechanisms. In J. J Laffont, editor, *Aggregation and Revelation of Preferences*. 1979.
12. H. Moulin. Almost budget-balanced vcg mechanisms to assign multiple objects. *Journal of Economic Theory*, 2008. In Press.
13. W. Vickrey. Counterspeculation, auctions, and competitive sealed tenders. *Journal of Finance*, 16(1):8–37, March 1961.

## A   Ordering of the Agents Based on Bid Profiles

We will define a ranking among the agents. This ranking is used crucially in proving a Theorem 1 on rebate function. This theorem is similar to Cavallo's theorem on characterization of DSIC, deterministic, anonymous rebate functions for homogeneous objects. We would not be actually computing the order among the bidders. We will use this order for proving impossibility of the linear rebate function with the desired properties.

### A.1   Properties of the Ranking System

When we are defining ranking/ordering among the agents, we expect the following properties to hold true:

– Any permutation of the objects and the corresponding permutation on bid vector, $(b_{i_1}, b_{i_2}, \ldots, b_{i_p})$ for each agent $i$, should not change the ranking. That is, the ranking should be independent of the order in which the agents are expected to bid for this objects.
– Two bidders with the same bid vectors should have the same rank.
– By increasing the bid on any of the objects, the rank of an agent should not decrease.

### A.2   Ranking among the Agents

This is a very crucial step. First, find out all feasible allocations of the $p$ objects among the $n$ agents, each agent receiving at most one object. Sort these allocations, according to the valuation of an allocation. Call this list $\mathcal{L}$. To find the ranking between $i$ and $j$, we uses the following algorithm.

1. $\mathcal{L}_{ij} = \mathcal{L}$
2. Delete all the allocations from $\mathcal{L}_{ij}$ which contain both $i$ and $j$.
3. Find out the first allocation in $\mathcal{L}_{ij}$ which contains one of the agent $i$ or $j$. Say $k'$.
   (a) Suppose this allocation contains $i$ and has value strictly greater than any of remaining allocations from $\mathcal{L}_{ij}$ containing $j$, then we say, $i \succ j$.
   (b) Suppose this allocation contains $j$ and has value strictly greater than any of remaining allocations from $\mathcal{L}_{ij}$ containing $i$, then we say, $j \succ i$.
4. If the above step is not able to decide the ordering between $i$ and $j$, let $\mathcal{A} = \{k \in K | v(k) = v(k')\}$. Update $\mathcal{L}_{ij} = \mathcal{L}_{ij} \setminus \mathcal{A}$ and recur to step (2) till EITHER
   • there is no allocation containing the agent $i$ or $j$ OR
   • the ordering between $i$ and $j$ is decided.
5. If the above steps do not give either of $i \succ j$ or $j \succ i$, we say, $i \equiv j$ or $i \succeq j$ as well as $j \succeq i$

Before we state some properties of this ranking system $\succeq$, we will explain it with an example. Let there be two items A and B, and four bidders. That is, $p = 2, n = 4$ and let their bids be: $b_1 = (4, 5), b_2 = (2, 1), b_3 = (1, 4)$, and $b_4 = (1, 0)$.

Now, allocation $(A = 1, B = 3)$ has the highest valuation among all the allocations. So,

$$\text{agent } 1 \succ \text{ agent } 2$$
$$\text{agent } 1 \succ \text{ agent } 4$$
$$\text{agent } 3 \succ \text{ agent } 2$$
$$\text{agent } 3 \succ \text{ agent } 4$$

Now, in $\mathcal{L}_{13}$ defined in the procedure above, the allocation $(A = 2, B = 1)$ has strictly higher value than any other allocation in which the agent 3 is present. So,

$$\text{agent } 1 \succ \text{ agent } 3.$$

Thus,

$$\text{agent } 1 \succ \text{ agent } 3 \succ \text{ agent } 2 \text{ and}$$

$$\text{agent } 1 \succ \text{ agent } 3 \succ \text{ agent } 4$$

In $\mathcal{L}_{24}$, the allocation $(A = 2, B = 1)$ has strictly higher value than any other allocation in which the agent 4 is present. Thus, the ranking of the agents is,

$$\text{agent } 1 \succ \text{ agent } 3 \succ \text{ agent } 2 \succ \text{ agent } 4$$

We can show that the ranking defined above, satisfies the following properties.

1. $\succcurlyeq$ defines a total order on set of bids.
2. $\succcurlyeq$ is independent of the order of the objects.
3. If two bids are the same, then they are equivalent in this order.
4. By increasing a bid, no agent will decrease his rank.

If agent $i \succcurlyeq$ agent $j$, we will also say $v_i \succcurlyeq v_j$.